# Depinning of the Charge-Density Waves in Quasi-2D 1T-TaS$_2$ Devices Operating at Room Temperature

**A. Mohammadzadeh[1,×], A. Rehman[2,×], F. Kargar[1,3], S. Rumyantsev[2], J. M. Smulko[4], W. Knap[2,5,6] R. K. Lake[7] and A.A. Balandin[1,3,*]**

[1]Nano-Device Laboratory, Department of Electrical and Computer Engineering, Bourns College of Engineering, University of California, Riverside, California 92521 USA

[2]CENTERA Laboratories, Institute of High-Pressure Physics, Polish Academy of Sciences, Warsaw 01-142 Poland

[3]Phonon Optimized Engineered Materials Center, Materials Science and Engineering Program, Bourns College of Engineering, University of California – Riverside, California 92521 USA

[4]Department of Metrology and Optoelectronics, Gdańsk University of Technology, Gdańsk 80-233 Poland

[5]Centre for Advanced Materials and Technologies CEZAMAT, Warsaw University of Technology, Warsaw 02-822 Poland

[6]Laboratoire Charles Coulomb, University of Montpellier and CNRS, Montpellier, 34950 France

[7]Laboratory for Terascale and Terahertz Electronics, Department of Electrical and Computer Engineering, University of California, Riverside, California 92521 USA

× Contributed equally to this work.
* Corresponding author (A.A.B.): balandin@ece.ucr.edu ; web-site: http://balandingroup.ucr.edu/

**Abstract**

We report on depinning of nearly-commensurate charge-density waves in 1T-TaS$_2$ thin-films at room temperature. A combination of the differential current-voltage measurements with the low-frequency noise spectroscopy provide unambiguous means for detecting the depinning threshold field in quasi-2D materials. The depinning process in 1T-TaS$_2$ is not accompanied by an observable abrupt increase in electric current – in striking contrast to depinning in the conventional charge-density-wave materials with quasi-1D crystal structure. We explained it by the fact that the current density from the charge-density waves in the 1T-TaS$_2$ devices is orders of magnitude smaller than the current density of the free carriers available in the discommensuration network surrounding the commensurate charge-density-wave islands. The depinning fields in 1T-TaS$_2$ thin-film devices are several orders of magnitude larger than those in quasi-1D van der Waals materials. Obtained results are important for the proposed applications of the charge-density-wave devices in electronics.

**Keywords:** charge-density-waves; depinning; noise spectroscopy; 2D van der Waals materials

Recently, charge-density-wave (CDW) materials and devices attracted renewed interest in the context of two-dimensional (2D) van der Waals materials.[1–6] The 1T polymorph of TaS₂ is one of 2D van der Waals materials of the transition-metal dichalcogenide (TMD) group that reveals several CDW phase transitions in the form of resistivity changes and hysteresis. The transitions can be induced by temperature, electric bias and other stimuli.[7–13] Two of the phase transitions in 1T-TaS₂ are above room temperature (RT). The CDW phase is a periodic modulation of the electronic charge density, accompanied by distortions in the underlying crystal lattice, and it may be commensurate, nearly commensurate, or incommensurate with the underlying crystal lattice.[14,15] Below ~180 K – 200 K, 1T-TaS₂ is in the commensurate CDW (C-CDW) phase. In this phase, the lattice reconstructs in the basal plane. The atomic positions slightly shift in a periodic star-of-David pattern forming a new 2D hexagonal supercell containing 13 Ta atoms with supercell lattice constants $a_{CDW} = \sqrt{13}a_0$. Theory has predicted and scanning tunneling microscopy (STM) measurements have shown this phase to be a Mott insulator state with one unpaired d-electron in each $\sqrt{13} \times \sqrt{13}$ superlattice cell.[16] Above ~200 K, the C-CDW phase undergoes transition to the nearly-commensurate CDW (NC-CDW) phase. This phase consists of islands of coherent C-CDW phase separated by a discommensurate network. The discommensurate regions can be viewed as domain walls that separate the ordered C-CDW islands.[17] At approximately 350 K, the NC-CDW phase transitions to the incommensurate CDW (IC-CDW) phase in which the translation vectors describing the periodicity of the CDW are no longer integer multiples of the original lattice constants. Finally, at ~500 K – 600 K, the material becomes a normal metal. The phase transitions at ~200 K and 350 K are accompanied by a hysteresis in the current-voltage response. The hysteresis of the NC-IC transition at 350 K has been recently exploited in a number of demonstrated device applications.[4,18–23]

The incommensurate CDW has translational invariance with respect to the lattice, which means that in an ideal lattice, there is no energy barrier to translation. In practice, there are impurities or defects that pin the IC-CDW to the lattice so that a finite electric field, denoted as the threshold field, $E_T$, is required to depin the IC-CDW and cause it to start *sliding* across the lattice. The sliding of the IC-CDW results in a collective current with a non-linear current-voltage (I-V) response, and it is generally viewed as additive to the normal current carried by the thermally excited electrons. At low temperature, when the normal current is suppressed, the depinning of the CDW reveals itself *via* an abrupt turn-on of the current when the electric field due to the applied voltage reaches the threshold field. As the peaks of the sliding CDW reach the contact, they give rise to a periodic modulation of the current resulting in an AC component known as "narrow band noise" (NBN). The average fundamental frequency of the NBN is related to the velocity, $v_{CDW}$, and wavelength, $\lambda$, of the CDW as $\overline{f}_0 = v_{CDW}/\lambda$.

The IC-CDW phase is common in the quasi-1D materials, and CDW sliding has been observed in NbSe₃, NbS₃, TaS₃, (TaSe₄)I, A₀.₃MoO₃ where A={K, Rb, Tl}, (NbSe₄)₁₀I₃, and TTF-TCNQ.[24,25] Interestingly, it has been noted that among the many known CDW materials, the CDW sliding has only been observed in the small subset listed above.[25] The threshold fields for depinning of CDWs in the quasi-1D materials are extremely low, ranging from 40 mV/cm to 4 V/cm.[14] Since the commensurate and nearly-commensurate CDWs are locked to the lattice, there is a large energy barrier to translate the CDW across the lattice. Therefore, it was generally believed that sliding could only occur in the incommensurate phase. Nevertheless, in some materials, I-V measurements have shown depinning and sliding below the C-CDW transition temperature. Orthorhombic TaS₃ transitions to a C-CDW at ~100 K, and the threshold field remains finite but increases as the temperature is lowered below 100 K.[14] The C-CDW transition of blue bronze, K₀.₃MoO₃, also takes place at ~100 K, however, $E_T$ decreases as the temperature is lowered below 100 K.[26] Since the decrease in $E_T$ violates all expectations based on the conventional theories of sliding CDWs, other models have been considered such as depinned phase solitons.[27] In this sense, CDW depinning and sliding are not completely understood phenomena even in the conventional CDW materials with quasi-1D crystal structure.

The initial attempts to observe depinning and sliding of CDWs in quasi-2D met with difficulties owing to absence of a characteristic abrupt increase in current.[28] The NBN phenomenon, which is an AC signature of the CDW sliding in quasi-1D CDW materials, has not been observed in quasi-2D CDW materials. In our recent report on 1T-TaS₂, current characteristics similar to NBN were attributed to the NC-CDW – IC-CDW phase transition hysteresis rather than the CDW sliding.[29] The only study, which focuses on CDW depinning in quasi-2D materials, is limited to the layered DyTe₃ compound.[30] We have previously assigned certain peaks in the low-frequency noise spectral density to the CDW depinning in 1T-TaS₂.[12,31] However, we did not have a confirmation by other experimental methods, and could not explain the absence of the non-linear current component at the threshold field. Here we use a combination of the *differential* I-V characteristics and the low-frequency *noise spectroscopy* to unambiguously establish the depinning threshold in 1T-TaS₂, and elucidate its differences from that in conventional bulk CDW materials with quasi-1D crystal structure. The obtained results shed light on switching phenomena in quasi-2D CDW materials; and can help in developing novel functionalities for device applications.

For this study, we used high-quality single-crystal 1T-TaS₂ (2D Semiconductors) as the source material. Thin films of 1T-TaS₂ were mechanically exfoliated and placed on a Si/SiO₂ substrate utilizing an in-house built transfer system. The device structures were fabricated by the electron-beam lithography and lift-off of Ti/Au (20-nm/180-nm) deposited by the electron-beam evaporation. The I-V characteristics of two tested 1T-TaS₂ devices measured at RT are presented in Figure 1. The hysteresis in each I-V is a result of the of transition between the NC-CDW and IC-CDW phases in the 1T-TaS₂ device. The I-V characteristics are well reproducible and in agreement with literature reports.[3,7,9] The pronounced hysteresis, observed at the bias voltage $V \approx$ 1.25 V and $V \approx$ 1.8 V in the devices A and B, respectively, is due to the NC-CDW to IC-CDW phase transition. The presence of hysteresis allows one to use such I-Vs for designing voltage-controlled oscillators and other devices by adding a load resistor, *i.e.* negative feedback, even if the resistance change is not large.[4] The nature of the NC-CDW to IC-CDW phase transition in two-terminal 1T-TaS₂ devices on Si/SiO₂ substrate is thermal – the current in the channel raises the local temperature above 350 K due to Joule heating and induces the phase transition.[12,23] An optical image of a representative device structure and additional characterization data are provided in the Supplemental Materials.

[Figure 1 (a-b): Device microscopy and schematic; I-V characteristics]

Since at RT, 1T-TaS₂ is in the NC-CDW phase, which contains IC-CDW regions, the depinning of the NC-CDW can happen at any bias. We intentionally focus on I-V characteristics at small biases, away from the NC-CDW -- IC-CDW phase transition point, to identify signatures of possible CDW depinning and sliding. Figure 2 (a) shows the I-V characteristics at low bias, $V_D$<0.5 V. The I-Vs are almost *linear* under forward and reverse bias, without any signatures of super-linearity or hysteresis. For comparison, the inset shows the depinning and on-set of CDW sliding revealed by an *abrupt* increase in current in a conventional CDW material with a quasi-1D crystal structure.[14] However, the differential I-V characteristics, *dI/dV*, clearly indicate a threshold voltage $V_T \approx 0.15$ V at which the electron transport through 1T-TaS₂ abruptly changes (see Figure 2 (b)). This phenomenon has been reproducibly observed for several devices, and it is clearly seen for both forward and reverse voltage sweeps with some hysteresis (see Supplemental Materials). The corresponding threshold field, $E_T = V_T/L \sim 1$ kV/cm, is *two to four orders of magnitude* larger than the depinning threshold fields observed in the quasi-1D materials ($L \sim 1.5$ μm is the distance between the contacts). To further prove that fluctuations in *dI/dV* are the result of CDW depinning, we utilize low-frequency noise measurements, which revealed various phase transitions in quasi-2D CDW materials.[12,29,31,32]

[Figure 2 (a-b) I-V and differential characteristics]

The low-frequency noise, typically with the spectral density $S(f) \sim 1/f^{\gamma}$ (f is the frequency and parameter $\gamma \approx 1$), is found in almost all materials and devices.[33,34] The details of our noise measurement setup and procedures have been reported elsewhere.[12,31] Figure 3 (a) shows the normalized noise spectral density, $S_I/I^2$, as a function of frequency, *f*, for several bias voltages, selected near $V_T \approx 0.15$, where a sudden change in *dI/dV* has been observed. Overall, the noise spectral density follows the $1/f$ – type behavior, and increases as the bias voltage approaches $V_T$. In the vicinity of the depinning voltage, the noise spectra develop pronounced bulges (for example, see spectra at $V$=195 mV and $V$=200 mV in Figure 3 (a)). The noise evolution can be more readily observed in Figure 3 (b), which presents the normalized spectral density, $S_I/I^2$, as a function of the bias voltage, *V*, at fixed *f*=10 Hz. $S_I/I^2$ reproduces the hysteresis observed in *dI/dV*, which occurs at ~0.18 V, and ~0.15 V for the forward and reverse bias directions, respectively. The threshold voltages obtained from the noise measurements are in excellent agreement with those defined by the derivative of the I-V characteristics shown in Figure 2 (b).

[Figure 3 (a-b): Noise spectral density]

For a more detailed analysis, Figure 4 (a-b) shows the noise spectral density multiplied by the frequency, $S_I/I^2 \times f$, and the noise amplitude over a larger voltage and frequency range. From Figure 4 (a), one can clearly see the noise bulges for voltages above $V_T$. The noise bulges are reminiscent of the Lorentzian noise spectrum, which is described by the equation $S_I(f) = S_0 \times (f_c^2/(f^2 + f_c^2))$ where $f_c$ is the corner frequency of the spectrum defined as $f_c = (2\pi\tau)^{-1}$, $S_0$ is the frequency independent portion of the function $S_I(f)$ when $f < f_c$ and $\tau$ represents the characteristic time constant.[33,34] However, the attempts to fit the measured bulges demonstrated deviation from a Lorentzian shape, possibly due to presence of competing processes with several time constants resulting in inhomogeneous broadening (see Figure S3 (b) in the Supplemental Materials). Figure 4 (b) shows the noise amplitude, *i.e.* noise spectral density averaged over several frequencies as a function of the bias voltage. The noise amplitude changes over a *four orders of magnitude* at $V_T$. Such a large increase comes from both the emergence of the noise bulges and overall increase of the $1/f$ noise background. It is this exceptionally large increase in the noise spectral density, $S_I/I^2$, that allows for observation of the depinning in $dI/dV$ derivative (see Figure 2 (b)).

In conventional semiconductors, the noise bulges in the $1/f$ spectrum are often associated with the generation – recombination (G-R) noise, which results from the charge carrier trapping and de-trapping by one type of a dominant defect. However, the noise bulges, Lorentzian-shaped or otherwise, can also be related to phase transitions[12,31,32] and CDW sliding.[24] The common feature among various mechanisms leading to a Lorentzian noise spectrum is that it is produced by some process characterized only by one time constant. The presence of several time constants may result in the deviation from the Lorentzian shape as in our case. From the polynomial data fitting in Figure 4 (a) we determine that for the small current levels, there is almost an exact linear relation between the characteristic frequency, $f_c$, and the current (see inset to Figure 4 (b)). The time constants, $\tau_c = 2\pi/f_c$, for the noise bulge change from ~10 ms to 0.1 μs. The linear dependence of $f_c$ on the current in the channel of quasi-2D CDW material is consistent with the linear relationship between the frequency of the NBN, $\overline{f_0}$, and the CDW current, $I_{CDW}$, observed in the quasi-1D CDW materials.[14,15] These considerations suggest that the observed noise bulges, which emerge at the voltage $V = V_T$, are associated with the on-set of CDW propagation in quasi-2D material.

[Figure 4 (a-b): Lorentzian noise spectrum and noise amplitude]

To clarify the mechanism of the observed depinning, *i.e.* electric field induced *vs.* temperature, we consider the local heating. The temperature rise at the depinning point can be roughly estimated from the ratio of the bias voltage at the depinning and the NC-CDW – IC-CDW phase transition, $V_T/V_{NC-IC}$~0.1. The dissipation is proportional to $V^2$. It was previously established that in the considered quasi-2D 1T-TaS$_2$ devices the NC-CDW – IC-CDW phase transition is induced by the local heating by about ~50 K.[12,23] Taking the temperature increases near this transition as ~50 K, we estimate, from the $(V_T/V_{NC-IC})^2$ ratio, the temperature rise at the depinning point to be below ~1 K. A more elaborate calculation, considering the structure geometry and material parameters, performed using the model reported in Ref. 12 gives the temperature rise below 2.5 K. This means that the depinning is induced by the electric field rather than local temperature increase. We also note that the noise does not change around RT with the small increase in temperature. This supports our conclusion that in the quasi-2D 1T-TaS$_2$ devices, the depinning is a field-induced phenomenon unlike the NC-CDW – IC-CDW phase transition. The CDW depinning and sliding in the quasi-2D material is unusual in the sense that it does not give rise to an observed increase in the total current (see Figure 2 (a)). On a larger voltage scale (Figure 1), the current becomes super-linear above ~0.75 V corresponding to an electric field of 5 kV/cm. The different scale of the electric fields compared to those in the quasi-1D materials deserves further investigation.

In the quasi-1D materials, the presence of a CDW gaps the Fermi surface so that there are relatively few normal carriers. In contrast, in the NC-CDW phase of 1T-TaS$_2$, the C-CDW islands should most likely be Mott insulators consistent with the C-CDW phase while the discommensurate boundary regions are semi-metallic.[35] Thus, there are plenty of free carriers present, and the resistivity resulting from the free carriers is ~$1.5 \times 10^{-3}$ $\Omega$cm. In the simple model of a sliding CDW, the velocity is $v = \bar{f}\lambda$, where $\lambda$ is the CDW wavelength. Then $J_{CDW} = env$. In this case, there is one electron for each cell of the C-CDW 2D supercell. Let $\lambda$ be equal one edge of the 2D supercell, $\lambda = \sqrt{13}a_0$, where $a_0 = 3.365$ Å is the in-plane lattice constant in the normal phase. The area of the supercell is $\lambda^2 \frac{\sqrt{3}}{2} = 1.3$ nm$^2$. Thus, $n_{2D} = 7.8 \times 10^{13}$ cm$^{-2}$, and $n_{3D} = \frac{n_{2D}}{c_0} =$ $1.3 \times 10^{21}\ cm^{-3}$ where $c_0$ is the out-of-plane lattice constant. For the frequency $\bar{f} = 60$ kHz, corresponding to a voltage of 0.45 V, we obtain the velocity $v = \bar{f}\lambda = 7.3 \times 10^{-3}$ cm/s, and the CDW collective current density $J_{CDW} = env = 1.5$ A/cm$^2$. Since the frequency, and thus the velocity, is linear with the voltage, the resistivity corresponding to the CDW current is $\rho_{CDW} = \frac{(V-V_T)}{J_{CDW}}\frac{1}{L} = 1.3$ k$\Omega \cdot$ cm, which is *six* orders of magnitude larger than that due to the normal current. Thus, the current contribution due to the sliding CDW is too small to be observed by the current measurements. As a result, we have an interesting feature of CDW sliding in quasi-2D materials: the depinning does not result in the observable on-set of non-linear current. One can still observe the depinning in the noise spectra and *dI/dV* characteristics since even slowly moving CDW can increase the scattering of the normal electrons, *e.g.* via their backflow[24], screening or other mechanisms, which deserve a separate study. The non-linearity, which is observed at higher fields (see Figure 1) as the bias approaches the NC-CDW to IC-CDW transition is likely related to Joule heating.

We also consider the processes of sliding discommensurations, and stochastic generation of free electrons from the C-CDW islands. The electron density associated with the commensurate islands is $n_{3D} = \frac{1}{c_0\lambda^2}\frac{2}{\sqrt{3}} \approx \frac{1}{c_0\lambda^2}$. The velocity is $\bar{f}\lambda$. Therefore, the CDW current is $J_{CDW} = \frac{e\bar{f}}{c_0\lambda}$. If instead of using $\lambda$ of the C-CDW supercell, we use $\lambda' = m\lambda$ of the average NC-CDW island size, and picture the islands moving, or equivalently, the discommensurations (dcm) moving, then $v_{dcm} = \bar{f}\lambda' = m\bar{f}\lambda$. The charge density is still $\frac{1}{c_0\lambda^2}$, so the current only changes by a factor of $m$, and $m$ is approximately 2-3 at RT.[17] Therefore, the value for $J_{CDW} = 1.5$ A/cm$^2$ is a good order-of-magnitude estimate, if we assume that the bulge in the noise spectrum results from a movement of the c-CDW islands or the discommensurations, *i.e.* domain walls, separating the islands. Another perspective, is to view the noise as similar to the G-R noise with a time constant corresponding to the release of the bound electron from the CDW cell. Along the contact, any released electron will be swept into the contact. The 2D density of electrons in the sheet of CDW cells next to the contact is $n_s = \frac{1}{c_0\lambda}$, and the current would then be $J_{RG} = \frac{en_s}{\tau} = \frac{e\bar{f}}{c_0\lambda} = J_{CDW}$. Thus, we obtain the same estimate for the magnitude of the CDW current from either perspective.

In conclusion, we investigated depinning of CDWs in 1T-TaS$_2$ thin-film devices operating in the NC-CDW phase. A combination of the differential I-V measurements with the low-frequency noise spectroscopy provided unambiguous means for detecting the depinning threshold field, $E_T$, in quasi-2D materials. The depinning process in the NC-CDW phase of 1T-TaS$_2$ is not accompanied by any observable increase in the electric current. This is explained by the fact that the velocity and the current density of CDWs, determined from the 60 kHz characteristic frequency, are low, $\sim 70\ \mu m/s$ and $\sim 1.5$ A/cm$^2$, and the depinning field is high, $E_T \approx 1$ kV/cm. The large electric field acting on the semi-metallic electrons in the discommensuration network results in a large normal current $\sim 2$ MA/cm$^2$ that obscures the CDW current. The on-set of non-linearity in the measured current at the biases larger than the depinning bias, $V_T$, is attributed to self-heating rather than the CDW current. The depinning field is two to four orders of magnitude higher than those in the quasi-1D van der Waals materials. Our results demonstrate substantial differences between depinning in quasi-2D and quasi-1D CDW van der Waals materials.

## Acknowledgements

The work at UC Riverside was supported, in part, by the U.S. Department of Energy Office of Basic Energy Sciences under the contract No. DE-SC0021020 "Physical Mechanisms and Electric-Bias Control of Phase Transitions in Quasi-2D Charge-Density-Wave Quantum Materials". A.R., W.K., and S.R. acknowledge the support from the CENTERA Laboratories in the framework of the International Research Agendas Program for the Foundation for Polish Sciences co-financed by the European Union under the European Regional Development Fund (No. MAB/2018/9). J.M.S acknowledges the partial support by the National Science Centre, Poland, the research project: 2019/35/B/ST7/02370. The fabrication of the CDW devices used in this study was performed at the UCR Nanofabrication Facility.

## Author Contributions

A.A.B. conceived the idea, coordinated the project, contributed to experimental data analysis, and wrote the initial draft of the manuscript; A.M. fabricated devices, conducted I-V measurements, and analyzed experimental data; A.R. performed the low-frequency noise measurements; S.L.R., F.K., J.M.S. and W.K. contributed to the experimental data analysis. R.K.L. led the theoretical analysis. All authors contributed to the manuscript preparation.

## Supplemental Information

The supplemental information is available at the Applied Physics Letters journal web-site for free of charge.

## The Data Availability Statement

The data that support the findings of this study are available from the corresponding author upon reasonable request.

## FIGURE CAPTIONS

**Figure 1:** Current-voltage characteristic of the 1T-TaS₂ devices on Si/SiO2 substrate at room temperature. The direction of the current sweep is indicated with the arrows. The data are presented for two devices with different channel length fabricated on the same structure. The arrows indicate the direction of the current sweep. The observed hysteresis is due to the transition between NC-CDW and IC-CDW phases induced by the channel current. The dashed straight lines are shown for comparison to visualize the gradual evolution of the current-voltage characteristics from linear to super-linear. The inset shows the design of the two-terminal 1T-TaS₂ devices used in this study.

**Figure 2.** (a) Close-up of the current-voltage characteristics of 1T-TaS₂ device A for the small bias voltage range, $V_D$<0.5 V. Notice that the current in the forward (red) and reverse (blue) sweeping overlaps. The straight black line is shown for comparison. No deviations from the non-

linearity are observed in this bias range. (b) The derivative of current-voltage characteristics revealing a strong change in the electron transport. The fluctuations in $dI/dV(V_D)$ are attributed to the depinning and on-set of CDW sliding.

**Figure 3:** (a) Normalized current noise spectral density, $S_I/I^2$, as a function of frequency at different applied bias voltages. Note a strong increase in the noise level as the bias voltage approaches the depinning point. The noise spectrum evolves from 1/f to develop the Lorentzian spectral features. (b) The normalized noise spectral density as a function of the bias voltage at a fixed frequency $f$=10 Hz. The noise reveals a hysteresis, which indicates the depinning points for the forward and reverse bias.

**Figure 4:** (a) Normalized current noise spectral density multiplied by the frequency, $S_I/I^2 \times f$, as a function of frequency at different applied bias voltages. The noise bulges and their evolution with the increasing current are clearly visible. (b) The noise amplitude as a function of the bias voltage. Note the break in the y-axis. The noise level experiences a drastic increase at the depinning point. The inset shows the dependence of the corner frequencies extracted from panel (a) with the current in the device channel.

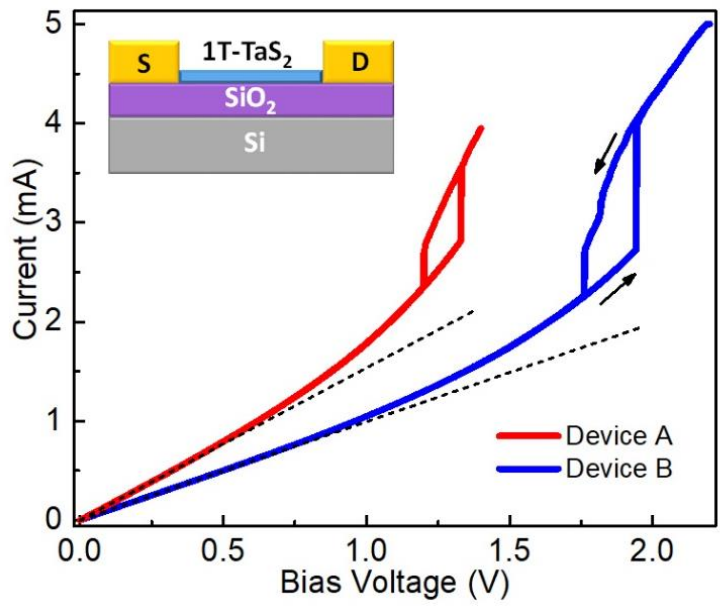

Figure 1

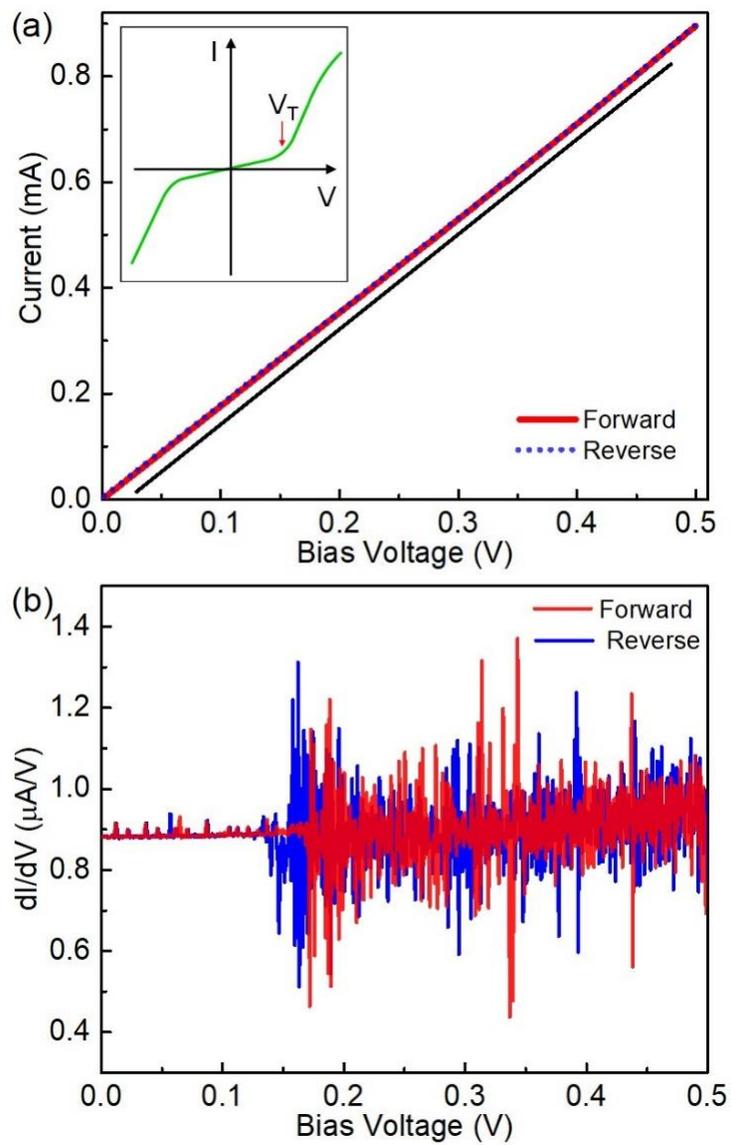

Figure 2

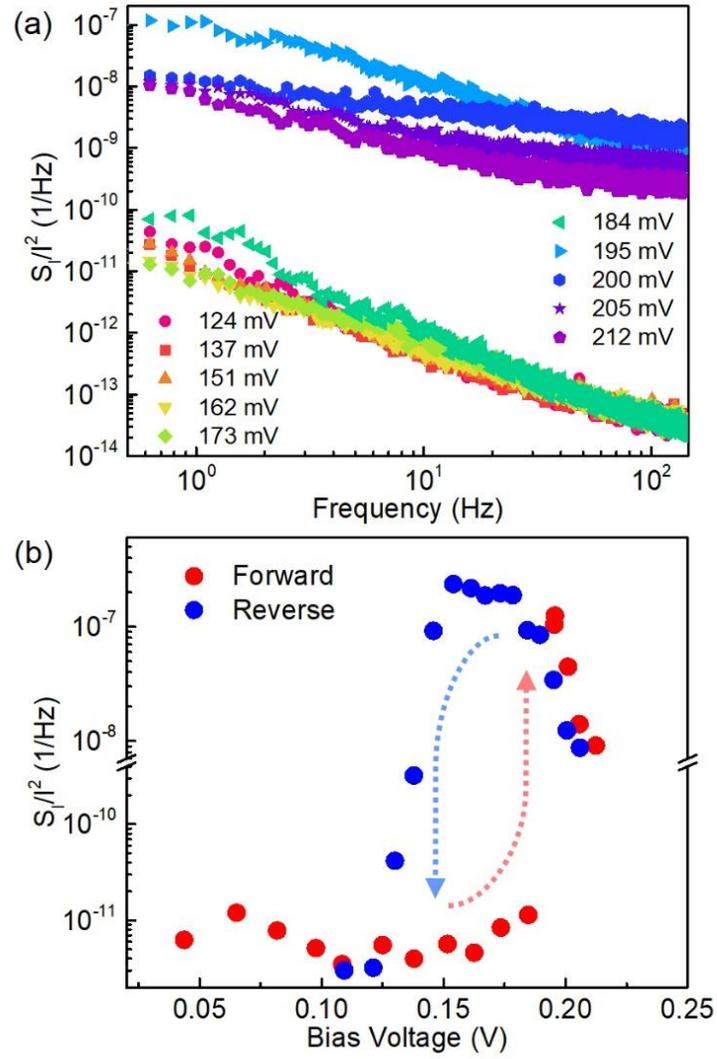

Figure 3

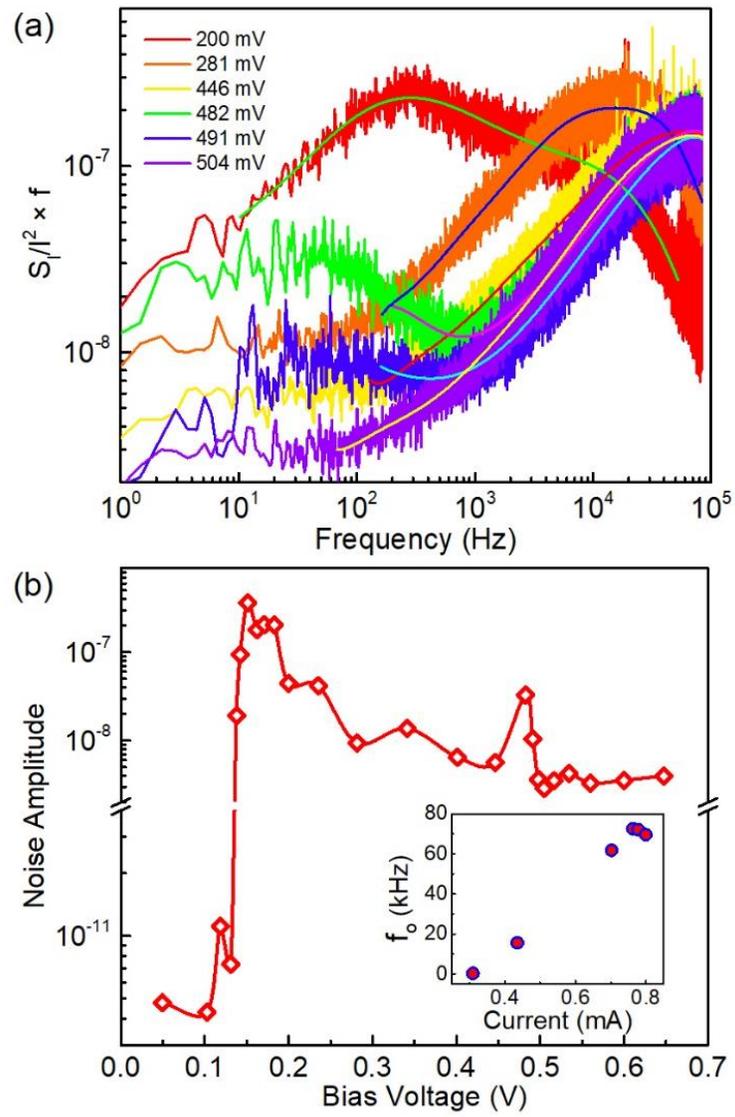

Figure 4